\documentclass[aip,jcp,amsmath,amssymb,preprint]{revtex4-1}
\usepackage[utf8]{inputenc} 

\usepackage{amsmath,amssymb}
\usepackage{graphicx} 
\usepackage{color}

\usepackage{booktabs} 
\usepackage{array} 
\usepackage{paralist} 
\usepackage{verbatim} 
\usepackage{subfig} 
\usepackage{hyperref}

\makeatletter
\newcommand*{\rom}[1]{\expandafter\@slowromancap\romannumeral #1@}
\makeatother

\renewcommand{\AA}{\mathcal{A}}
\newcommand{\CC}{\mathcal{C}}
\newcommand{\LL}{\mathcal{L}}
\newcommand{\dd}{\mathrm{d}}

\newcommand{\ii}{\text{i}}

\newcommand{\av}[1]{\langle #1 \rangle}
\newcommand{\avg}[2]{\langle #1 \rangle_{#2}}
\newcommand{\expo}[1]{\text{exp}\left( #1 \right)}
\newcommand{\e}{\text{e}}
\newcommand{\tS}{\text{S}}
\newcommand{\tL}{\text{L}}
\newcommand{\tR}{\text{R}}

\newcommand{\tB}{\text{B}}
\newcommand{\tSB}{\text{SB}}

\usepackage{braket}

\begin{document}
\title{Universal approach to quantum thermodynamics of strongly coupled systems under nonequilibrium conditions and external driving}
\author{Wenjie Dou}
\email{douw@berkeley.edu}
\thanks{These authors contributed equally}
\affiliation{Department of Chemistry, University of California Berkeley, Berkeley, California 94720, United State} 

\author{Jakob B\"atge}
\email{jakob.baetge@physik.uni-freiburg.de}
\thanks{These authors contributed equally}
\affiliation{Institute of Physics, Albert-Ludwigs University Freiburg, Hermann-Herder-Str. 3, 79104 Freiburg, Germany}

\author{Amikam Levy} 
\email{amikamlevy@gmail.com}
\affiliation{Department of Chemistry, University of California Berkeley, Berkeley, California 94720, United State}
\affiliation{The Raymond and Beverly Sackler Center for Computational Molecular and Materials Science, Tel Aviv University, Tel Aviv, Israel 69978}
\author{Michael Thoss}
\email{michael.thoss@physik.uni-freiburg.de}
\affiliation{Institute of Physics, Albert-Ludwigs University Freiburg, Hermann-Herder-Str. 3, 79104 Freiburg, Germany}

\begin{abstract}

We present an approach based on {a} density matrix expansion to study thermodynamic {properties} of a quantum system strongly coupled to two or more baths. For slow external driving of the system, we identify the adiabatic and nonadiabatic contributions to thermodynamic quantities, and we show how the first and second laws of thermodynamics are manifested in the strong coupling regime. Particularly, we show that  the entropy production is positive up to second order in the driving speed. {The formulation can be applied both for Bosonic and Fermionic systems,} and recovers  previous results for the equilibrium case (Phys. Rev. B 98, 134306 [2018]). {The approach is then demonstrated for the driven resonant level model as well as the driven Anderson impurity model, where the hierarchical quantum master equation method is used to accurately simulate the nonequilibrium quantum dynamics.} 

\end{abstract}
 
\maketitle

\section{Introduction}

Thanks to the advance in nanofabrication, quantum information and computing technologies, there has been an increasing research interest in the study of dynamics and thermodynamics for small systems consisting of just a few atoms (or photons, spins, etc.). Being far from the thermodynamic limit, these systems are subject to strong fluctuations and/or are not necessarily weakly coupled to their environments in general. Hence, the concepts of quantum thermodynamics emerge, addressing the quantum nature of thermodynamic quantities \cite{gemmer2009quantum,kosloff2013quantum, PhysRevLett.114.080602, vinjanampathy2016quantum,anders2017focus,PhysRevLett.116.240403,alicki2018introduction,benenti2017fundamental,RevModPhys.83.771,PhysRevLett.102.210401}, e.g. entropy production, dissipation and fluctuation, energy flow and work efficiency.\cite{mishaPRBfriction,millen2016perspective, gemmer2009quantum, talkner2007fluctuation,brandao2015second,jarzynski2011equalities} While theoretical works have focused on formulating thermodynamic laws for quantum systems, recent experiments have started to test the concepts of quantum heat engines. \cite{poot2012mechanical, pekola2015towards, rossnagel2016single,PhysRevE.96.052106,PhysRevLett.122.110601}

Different from the regimes of weak system-bath coupling, where quantum thermodynamics have been successfully formulated for certain systems \cite{spohn1978irreversible,RevModPhys.81.1665,kosloff2013quantum,gelbwaser2015thermodynamics,PhysRevLett.119.050601,PhysRevE.85.061126,kosloff2014quantum}, the regimes of strong system-bath coupling remain as open questions. On the one hand, there have been plenty studies focusing on the dynamics and transport properties of strongly coupled nano systems, using either numerically exact methods (e.g. multilayer multiconfiguration time dependent Hartree (ML-MCTDH) \cite{wang2009numerically,wang2013multilayer,wang2018multilayer}, path integral and quantum Monte Carlo \cite{prldata,PhysRevB.96.155126,segal2010numerically,werner2009diagrammatic}, the hierarchical quantum master equation (HQME) \cite{tanimura2006stochastic,ThossHEOM,erpenbeck2019hierarchical,jin2008exact}) or approximate approaches (e.g. numerical renormalization group \cite{anders2008steady,heidrich2009real}, combinations of reduced density matrix techniques and impurity solvers \cite{cohen2011memory,wilner2015sub,kidon2018memory},  nonequilibrium Green's function \cite{inelastic,mishaPRBfriction,PhysRevLett.114.080602,erpenbeck2015effect,hartle2008multimode}, scattering theory \cite{vonOppenPRB,bruch2018landauer,beilstein}, and mapping techniques \cite{strasberg2016nonequilibrium,PhysRevE.95.032139,PhysRevB.97.205405,gelbwaser2015strongly,katz2016quantum}). On the other hand, the thermodynamic properties of these systems are less understood, particularly in the case with external driving. One main challenge in the study of thermodynamics of strongly coupled nano systems is how to properly quantify energy, heat, and entropy for a system which is strongly hybridized with baths. Another way of looking at the problem is how to treat/split the interactions between the system and baths. \cite{solinas2013work,schmidt2015work,gogolin2016equilibration,subacsi2012equilibrium,ness2017nonequilibrium,PhysRevX.7.021003}

As an example for a noninteracting nano  system, the driven resonant level model has been studied  extensively in the literature. \cite{mishaPRBfriction,PhysRevLett.114.080602,PhysRevB.93.115318} Within the wide-band approximation, studies have shown that a symmetric splitting of the interactions between the system and bath is able to describe thermodynamic quantities for the extended system consistently. \cite{PhysRevB.93.115318} To the second order in driving speed, the entropy production is positive, and is related to dissipated work (i.e. frictional effects) at equilibrium. \cite{PhysRevLett.114.080602,PhysRevB.93.115318} Similar results have been obtained for bosonic systems. \cite{PhysRevB.94.035420} However, such a symmetric splitting may not be able to describe higher moments in thermodynamic quantities correctly. Later, von Oppen and co-workers employed the concept of scattering states to avoid the splitting of the system-bath couplings. \cite{bruch2018landauer} {Nevertheless}, their approach as well as most studies of quantum thermodynamics in the strong coupling limits are restricted to noninteracting systems and equilibrium cases.

In a recent publication \cite{PhysRevB.98.134306}, one of the authors and co-workers  have proposed a generic approach to study quantum thermodynamics at equilibrium. The approach is based on a description of the full density matrix (including system and bath). Under slow external driving, the full density matrix is expanded into a series of terms in the power of driving speeds, where the adiabatic and non-adiabatic contributions to the thermodynamic quantities are identified, and further, the first and second law of thermodynamics are formulated. The entropy production rate is found to be positive, and is related to dissipative work. This general formulation can be applied to interacting systems as well. When strong electron-electron (el-el) interactions are allowed in the Anderson impurity model, Kondo signatures are found in thermodynamic quantities, e.g. nonadiabatic energy, dissipated work. \cite{PhysRevLett.119.046001}

In the present work, we extend the previous study based on nonadiabatic expansion of the full density matrix to the nonequilibrium case. 
{Nonequilibrium conditions can be achieved by having the subsystem coupled to two (or more) baths that induce energy flows due to different temperatures and/or chemical potentials.} 
When subject to external driving, we establish a thermodynamic description for the case of finite driving speeds. In addition, we show that the nonadiabatic entropy production rate can be recasted into a Kubo transformed correlation function and remains positive, such that the second law of thermodynamics holds out of equilibrium. 
{We apply our analysis to the resonant level model as well as the Anderson impurity model, and further study thermodynamic signatures arising from el-el interactions.}

The paper is organized as follows. In Sec. \ref{sec:adiabatic}, we formulate thermodynamic laws {for out-of-equilibrium systems within the adiabatic limit.} In Sec. \ref{sec:nonadiabatic}, {we extend the results to} the nonadiabatic limit and identify the entropy production rate.  In Sec. \ref{sec:model}, we apply our analysis as well as numerical simulations using the HQME method to model systems. Finally, we conclude in Sec. \ref{sec:conclusion}.

\section{adiabatic thermodynamics} \label{sec:adiabatic}
In this section, we consider thermodynamics for a nonequilibrium system under infinitely slow driving, i.e. in the adiabatic limit. The thermodynamic quantities can be defined using the steady state density matrix.

\subsection{Steady state solution of an undriven system}
\begin{figure}[htbp] 
   \centering
   \includegraphics[width=4in]{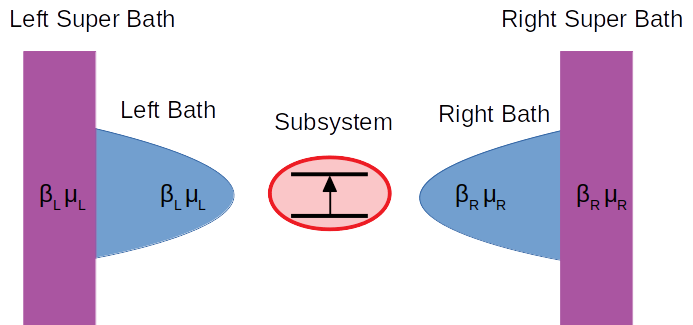} 
   \caption{A sketch of an out-of-equilibrium system. A subsystem is strongly coupled to multiple baths with different temperatures and chemical potentials. The super baths are  weakly coupled to the baths to make sure that the total system (subsystem+baths) reaches steady state. The subsystem can be subject to external driving.}
   \label{fig:sketch}
\end{figure}

We first consider thermodynamic {properties of} a nonequilibrium quantum system in the the static limit, i.e. without external driving. We assume the dynamics of the total system, including a subsystem and multiple baths, are governed by the total Hamiltonian $\hat H$. The total density matrix $\hat \rho$  follows the Liouville equation, 
\begin{eqnarray}
\frac{\partial}{\partial t} \hat \rho = -\frac{i}{\hbar} [ \hat H, \hat \rho ] .
\end{eqnarray}
{To mimic the steady state solution of the full density matrix (baths+subsystem) we introduce super baths  as illustrated in Fig. \ref{fig:sketch}. The steady state full density matrix satisfies}  
\begin{eqnarray}
\partial_t \hat \rho_{ss} =  -\frac{i}{\hbar} [ \hat H, \hat \rho_{ss} ] = 0 .
\end{eqnarray}
Note that, in the equilibrium case, the total system maintains one temperature ($k_B T=1/\beta$) and one chemical potential $\mu$ due to weak coupling to a super bath (see also Ref.\ \onlinecite{PhysRevB.98.134306} for a discussion),  such that the equilibrium solution to the density matrix is given by the Boltzmann/Gibbs distribution
\begin{eqnarray}
\hat \rho_{eq} =  e^{- \beta (\hat H-\mu \hat N )}/Z .
\end{eqnarray}
Here $\hat N$ is the number operator, and $Z = Tr(e^{- \beta (\hat H-\mu \hat N )})$ is the partition function. 

Under nonequilibrium conditions, where a subsystem couples to multiple baths with different temperatures or/and chemical potentials, the steady state density matrix does not admit a simple solution. Nevertheless, as shown by Hershfield\cite{PhysRevLett.70.2134} and others \cite{PhysRev.115.1405,ness2017nonequilibrium, zubarev1994nonequilibrium}, the steady-state density matrix can be formally expressed as
\begin{eqnarray} \label{eq:rho_ss}
\hat \rho_{ss} =  e^{-\bar{\beta} (\hat H- \hat Y )}/\Omega. 
\end{eqnarray}
Here $\Omega = Tr(e^{-\bar{\beta} (\hat H- \hat Y )})$ is the normalization factor and $k_B \bar T = 1/\bar{\beta}$ is the reduced temperature (e.g. for a subsystem coupled to two baths with inverse temperature $\beta_L$ and $\beta_R$, $\bar \beta = (\beta_L + \beta_R)/2$). $\hat Y$ in the above equation is an operator that accounts for particle transport throughout the subsystem. The formal expression for $\hat Y$ can be found in Ref.\ \onlinecite{e19040158} and can be obtained analytically for certain noninteracting cases (see Sec. \ref{sec:model}).    

With such a formal solution, we can define the steady state energy and entropy of the total system (baths + subsystem) as 
\begin{eqnarray}
E^{(0)} = Tr (\hat H \hat \rho_{ss} ), \\
S^{(0)} =  - k_B  Tr (\hat \rho_{ss} \ln \hat \rho_{ss} ) .
\label{eq:S_ss}
\end{eqnarray}
We use superscript $^{(n)}$ to indicate that the thermodynamic quantities are $n$th order in the driving
speeds. Here, the superscript $^{(0)}$ indicates that the quantities in the above equations are zeroth order in the driving
speeds (see below). 

\subsection{Adiabatic limit for a driven system}
In order to construct a heat engine or a refrigerator, we introduce additional external driving of our nonequilibrium system using a time-dependent Hamiltonian. Without loss of generality, we assume that the system Hamiltonian depends on a set of parameters,  i.e. $\hat H = \hat H(\bold R)$, and the parameters $\bold R = (R_1, R_2, ...,R_{\alpha},... )$ vary in time due to external driving. 

In the adiabatic limit, where the driving speed is very small as compared to the system dynamics, i.e. $\dot{\bold R} \approx 0$,   the system remains at steady state and follows the instantaneous Hamiltonian. By taking the time derivative, we can define the rate of change of thermodynamic quantities. Particularly,  the rate of change of the total energy is (note that the steady state density matrix also depends on $\bold R$):
\begin{eqnarray} \label{eq:E(1)}
\dot E^{(1)} = \sum_{\alpha} \dot R_\alpha  \partial_\alpha E^{(0)} = \sum_{\alpha} \dot R_\alpha  Tr(\hat \rho_{ss} \partial_\alpha \hat H)  + \sum_{\alpha} \dot R_\alpha  Tr(\hat H \partial_\alpha  \hat \rho_{ss}).
\end{eqnarray}
Here we have denoted $ \partial_\alpha \equiv \frac{\partial}{\partial R_\alpha}$. The superscript $^{(1)}$ indicates that the quantities are first order in the driving speeds.

Naturally, we can define the rate of heat transport
\begin{eqnarray} \label{eq:Q(1)}
\dot Q^{(1)} = \sum_{\alpha} \dot R_\alpha Tr ((\hat H -\hat Y) \partial_\alpha  \hat \rho_{ss} ),
\end{eqnarray}
the rate of work done to the system
\begin{eqnarray} \label{eq:W(1)}
\dot W^{(1)} = \sum_{\alpha} \dot R_\alpha Tr (\partial_\alpha  \hat H \hat \rho_{ss} ),
\end{eqnarray}
as well as the rate of change in energy due to particle transport 
\begin{eqnarray} \label{eq:Y(1)}
\dot Y^{(1)} = \sum_{\alpha} \dot R_\alpha Tr (\hat Y \partial_\alpha  \hat \rho_{ss} ).
\end{eqnarray}
With such definitions, we note that the rate of change in energy $\dot E^{(1)}$ is equal to the combination of  the rate of heat transport $\dot Q^{(1)}$,  the rate of work $\dot W^{(1)}$, and the rate of change in energy due to particle transport $\dot Y^{(1)}$,
\begin{eqnarray}
\dot E^{(1)}  = \dot Q^{(1)}  + \dot W^{(1)} + \dot Y^{(1)}, 
\end{eqnarray}
showing that the first law of thermodynamics holds in the adiabatic limit. 

In addition, in the adiabatic limit, using the definitions in Eqs. (\ref{eq:rho_ss}), (\ref{eq:S_ss}) and (\ref{eq:Q(1)}),  we find that the rate of change in entropy is equal to the rate of change of heat:  
\begin{eqnarray}
\dot S^{(1)} = \sum_{\alpha} \dot R_\alpha \partial_\alpha S^{(0)} =  k_B \bar \beta \sum_{\alpha} \dot R_\alpha Tr ((\hat H -\hat Y) \partial_\alpha  \hat \rho_{ss} ) =\frac{\dot Q^{(1)}}{\bar T} .
\end{eqnarray}

At this point, a few words are appropriate regarding the definition of entropy and the heat flow as well as the meaning of super baths. Just as for the equilibrium case \cite{PhysRevB.98.134306}, in a closed system (subsystem+baths), the rate of change in heat flow and entropy will be zero even for the out-of-equilibrium case. To see this, take the rate of change in heat flow as an example (note that $\hat H$ and  $\hat Y$ commute \cite{ness2017nonequilibrium})
\begin{eqnarray}
\dot Q = Tr( (\hat H -\hat Y) \frac{ d \hat \rho }{dt} ) = -\frac{i}{\hbar} Tr( (\hat H -\hat Y)  [ \hat H, \hat \rho ] ) = 0.
\end{eqnarray}
By contrast, our definition of the rate of change in heat from Eq. (\ref{eq:Q(1)}) does not vanish. This apparent contradiction is due to the fact that we are not dealing with a closed system: The {presence} of the super baths guarantees a  unique steady state solution for  the total system. As a result, Eq. (\ref{eq:Q(1)}) defines the rate of heat exchange with the super baths. See also discussions in Ref.\ \onlinecite{PhysRevB.98.134306}. The same argument holds for entropy and energy. 

\section{nonadiabatic thermodynamics} \label{sec:nonadiabatic}
When the external driving is not infinitely slow as compared to the timescale of system relaxation, the total system does not necessarily remain at steady state,  
{hence} nonadiabatic effects arise. In this section, we quantify such nonadiabatic contributions to  
thermodynamic quantities and entropy production. 
\subsection{Expansion of density operator in driving speed} \label{subsec:expansion}
To systematically classify the nonadiabatic contributions, we use an expansion of the density operator in the driving speed. 
The procedure here follows Ref.\ \onlinecite{PhysRevB.98.134306}. To be self-consistent, we outline the main steps below.  

With finite driving speed, the equation of motion for the density matrix can be described as 
\begin{eqnarray} \label{eq:EOMfull}
\frac {d} {dt} \hat  \rho (\bold R, t) =  \frac {\partial } { \partial t} \hat  \rho  + \sum_\nu \dot{R}_\nu  \partial_\nu \hat \rho = - \frac{i}{\hbar} [ \hat H (\bold R), \hat  \rho] .
\end{eqnarray}
In presence of finite driving speed ($\dot{\bold R} \neq 0$), the total derivative respect to time $\frac {d} {dt}$ is a combination of the partial derivative respect to time $\frac {\partial } { \partial t} $ plus  driving terms $\sum_\nu \dot{R}_\nu  \partial_\nu$, i.e. $\frac {d} {dt} = \frac {\partial } { \partial t} + \sum_\nu \dot{R}_\nu  \partial_\nu$. 
Assuming that the driving speed $\dot{\bold R}$ is small,  the total density matrix can be then expressed as a series of terms in the order of the driving speed: 
\begin{eqnarray} \label{eq:fullRho}
\hat \rho  = \hat  \rho^{(0)} +\hat  \rho^{(1)} +\hat  \rho^{(2)} + \cdots
\end{eqnarray} 
Here $\hat \rho^{(n)}$ is density operator in $n$th order of $\dot{\bold R}$. We can break Eq. (\ref{eq:EOMfull}) into a series of equations by matching the order in the driving speed on both sides, 
\begin{eqnarray} \label{eq:EOM0}
\frac {\partial } { \partial t}  \hat  \rho^{(0)}  &=& - \frac{i}{\hbar} [ \hat H, \hat  \rho^{(0)} ], \\
\frac {\partial } { \partial t} \hat  \rho^{(n)}  &=& - \frac{i}{\hbar} [ \hat  H, \hat  \rho^{(n)} ]  - \sum_\nu \dot{R}_\nu  \partial_\nu \hat  \rho^{(n-1)}, \: n \ge 1.
\label{eq:EOMbig1}
\end{eqnarray}
Obviously, the steady state solution in Eq. (\ref{eq:rho_ss}) satisfies Eq. (\ref{eq:EOM0}), and we use the steady state density matrix as the zeroth order density operator, 
\begin{eqnarray}
\hat \rho^{(0)} = \hat \rho_{ss}. 
\end{eqnarray}
Starting with the zeroth order density matrix, we can then solve for the $n$th order $\hat \rho^{(n)}$ sequentially, 
\begin{eqnarray}
 \hat \rho^{(n)} (\bold R, t) &=&   - \sum_\nu    \int_0^t  e^{-i \hat  H (t-t')/\hbar} \dot{R}_\nu \partial_\nu \hat  \rho^{(n-1)} e^{i \hat H (t-t')/\hbar}  dt', \: n \ge 1.
\end{eqnarray}
If we assume that the timescale of bath relaxation is much faster than the speed of driving, we can invoke the Markovian approximation (i.e. the super baths bring the system back to steady state fast) in the above equation: 
\begin{eqnarray}
 \hat  \rho^{(n)} (\bold R) &\approx& - \sum_\nu \dot{R}_\nu  \int_0^\infty  e^{-i \hat H t'/\hbar} \partial_\nu \hat  \rho^{(n-1)} e^{i \hat  H t'/\hbar}  dt', \: n \ge 1.
\end{eqnarray}
Particularly, the first order correction to the steady state density matrix is 
\begin{eqnarray} \label{eq:rho_1}
 \hat  \rho^{(1)} (\bold R) &\approx& - \sum_\nu \dot{R}_\nu  \int_0^\infty  e^{-i \hat H t'/\hbar} \partial_\nu \hat  \rho_{ss} e^{i \hat  H t'/\hbar}  dt'. 
\end{eqnarray}

\subsection{Nonadiabatic corrections to thermodynamics quantities}
We now consider the rate of change in thermodynamic quantities to the second order in driving speed. This can be done by replacing the steady state solution in Eqs. (\ref{eq:E(1)})-(\ref{eq:Y(1)}) with the nonadiabatic correction, Eq. (\ref{eq:rho_1}).  We find that the nonadiabatic correction to the rate of change in energy is  
\begin{eqnarray}
\dot E^{(2)} = \sum_{\alpha} \dot R_\alpha  Tr(\hat \rho^{(1)} \partial_\alpha \hat H)  + \sum_{\alpha} \dot R_\alpha  Tr(\hat H \partial_\alpha  \hat \rho^{(1)}).
\end{eqnarray}
Correspondingly, the nonadiabatic correction to the rate of change in heat transport, work, and energy due to particle transport are given, respectively, by
\begin{eqnarray} \label{eq:Q(2)}
\dot Q^{(2)} = \sum_{\alpha} \dot R_\alpha Tr (( \hat H -\hat Y)  \partial_\alpha  \hat \rho^{(1)} ), \\
\dot W^{(2)} = \sum_{\alpha} \dot R_\alpha Tr (\partial_\alpha  \hat H \hat \rho^{(1)} ), \\
\dot Y^{(2)} = \sum_{\alpha} \dot R_\alpha Tr (\hat Y  \partial_\alpha  \hat \rho^{(1)} ) .
\end{eqnarray}
Consequently, the first law of thermodynamics holds in the nonadiabatic limit, 
\begin{eqnarray}
\dot E^{(2)}  = \dot Q^{(2)}  + \dot W^{(2)} + \dot Y^{(2)}.
\end{eqnarray}

We further note that the nonadiabatic correction to the rate of change in work is related to friction tensor, \cite{PhysRevB.96.104305,PhysRevB.97.064303,PhysRevLett.119.046001} 
\begin{eqnarray}
\dot W^{(2)} = \sum_{\alpha\nu} \dot R_\alpha \gamma_{\alpha \nu} \dot R_\nu,
\end{eqnarray}
where the friction tensor is defined as 
\begin{eqnarray}
 \gamma_{\alpha \nu}  =  \int_0^\infty  Tr ( e^{-i \hat H t'/\hbar} \partial_\nu \hat  \rho_{ss} e^{i \hat  H t'/\hbar} \partial_\alpha  \hat H )  dt'.
\end{eqnarray}
At equilibrium, due to time reversal symmetry, the friction tensor is symmetric (along with respect to $\alpha$ and $\nu$) and positive definite\cite{dou2018perspective, PhysRevB.96.104305,PhysRevLett.119.046001}, such that there is always a dissipated work associated with driving, i.e. $\dot W^{(2)} >0 $. Out of equilibrium, however, the presence of a nonequilibrium current can break the time reversal symmetry, such that the friction tensor is no longer symmetric nor positive definite.\cite{beilstein,lvPRBfriction} As shown by von Oppen \textit{et al}\cite{beilstein}, in a minimal setup of a two-level system with two external degrees of freedom ($\alpha$ and $\nu$), the negativity of the friction is present, for example, when an electron current pumps energy into the two-level system. 
 
\subsection{Entropy production and the second law of thermodynamics}

When the total system does not remain {at} steady state due to external driving, we define the total entropy using the total density matrix, such that 
\begin{eqnarray} \label{eq:fulS}
S =  - k_B  Tr (\hat \rho \ln \hat \rho ). 
\end{eqnarray}
To zeroth order in driving speed, the above definition recovers the steady state entropy in Eq. (\ref{eq:S_ss}). To first order in the driving speed, the entropy is then given by  
\begin{eqnarray} \label{eq:S1}
S^{(1)} =  - k_B  Tr (\hat \rho^{(1)} \ln \hat \rho_{ss} ),
\end{eqnarray}
as was shown in Ref.\ \onlinecite{PhysRevB.98.134306}.

The derivative of Eq. (\ref{eq:S1}) with respect to time gives the rate of change for the entropy to the second order in driving speed, 
\begin{eqnarray} \label{eq:Sad}
\dot S^{(2)} = \sum_{\alpha} \dot R_\alpha \partial_\alpha S^{(1)}  = -k_B \sum_{\alpha} \dot R_{\alpha} tr( \partial_\alpha \hat \rho^{(1)} \ln \hat \rho_{ss} ) -k_B \sum_{\alpha} \dot R_{\alpha} tr( \hat \rho^{(1)} \partial_\alpha \ln \hat \rho_{ss} ). 
\end{eqnarray}
Using the definition in Eq. (\ref{eq:Q(2)}), we note that the first term in the above equation is equal to $\frac{\dot Q^{(2)}}{\bar T}$, such that Eq. (\ref{eq:Sad}) can be rewritten as 
\begin{eqnarray} \label{eq:dotS2}
\dot S^{(2)}  = \frac{\dot Q^{(2)}}{\bar T}  + \Delta \dot S_{NA}, 
\end{eqnarray}
where we have defined 
\begin{eqnarray}
\Delta \dot S_{NA} = -k_B \sum_{\alpha} \dot R_{\alpha} tr(  \hat \rho^{(1)} \partial_\alpha \ln \hat \rho_{ss} ).
\end{eqnarray}
$\Delta \dot S_{NA}$ can be interpreted as the entropy production rate due to non-adiabatic driving. If we insert the result for $\hat \rho^{(1)}$, and use the following Campbell-Baker-Hausdorff formula,
\begin{eqnarray}
\partial_\nu \hat \rho_{ss} = \int_0^1 \hat \rho_{ss}^{1-\lambda} \partial_\nu(\ln \hat \rho_{ss}) \hat \rho_{ss}^{\lambda} d\lambda, 
\end{eqnarray}
$\Delta \dot S_{NA}$ can be rewritten as a Kubo transformed correlation function  
\begin{eqnarray} \label{eq:deltaS}
\Delta \dot S_{NA} =  k_B \sum_{\alpha \nu} \dot R_{\alpha} \dot R_{\nu}   \int_0^\infty  \langle \delta \hat{\mathcal{F}}_\alpha (t)  \delta \hat{\mathcal{F}}_\nu \rangle_{K} dt  > 0.
\end{eqnarray}
Here, we have defined the following operator in Heisenberg picture
\begin{eqnarray} 
\delta \hat{\mathcal{F}}_\alpha &=&   \partial_\alpha \ln \hat \rho_{ss}, \\
\delta \hat{\mathcal{F}}_\alpha (t) &=& e^{i \hat H t/\hbar}  \delta \hat{\mathcal{F}}_\alpha e^{-i \hat H t/\hbar} .
\end{eqnarray}
The Kubo transformed correlation function is given by
\begin{eqnarray}
\langle \delta \hat{\mathcal{F}}_\alpha (t)  \delta \hat{\mathcal{F}}_\nu \rangle_{K} =\int_0^1   Tr ( \hat \rho_{ss}^{1-\lambda}  \delta \hat{\mathcal{F}}_\nu \hat \rho_{ss}^{\lambda}  \delta \hat{\mathcal{F}}_\alpha (t) ) d\lambda.
\end{eqnarray}
Obviously, the Kubo transformed self-correlation function is positive definite such that the the entropy production rate is always positive, i.e. $\Delta \dot S_{NA} > 0$.  This can be shown using a Lehmann representation. Employing the eigenstates $|\Psi_n \rangle$ of the Hamiltonian, $\hat H |\Psi_n \rangle = E_n |\Psi_n \rangle$, and the steady state density operator, $\hat \rho_{ss} |\Psi_n \rangle = \rho_n |\Psi_n \rangle$, the entropy production rate in Eq. (\ref{eq:deltaS}) can be rewritten as
\begin{eqnarray}
\Delta \dot S_{NA} = k_B \sum_{mn} |\langle \Psi_n| \delta \hat{\mathcal{F}}| \Psi_m \rangle|^2  \delta(E_n - E_m)  \int_0^1 \rho_n^\lambda \rho_m^{1-\lambda} d\lambda .
\end{eqnarray}
Here, we have defined $\delta \hat{\mathcal{F}} = \sum_{\alpha} \dot R_{\alpha} \delta\hat{\mathcal{F}}_\alpha $. Note that every single term in the above equation is positive, such that entropy production rate $\Delta \dot S_{NA}$ is positive, i.e. the second law of thermodynamics holds.

Eqs. (\ref{eq:dotS2}) and (\ref{eq:deltaS}) are our main results. 
To better understand the entropy production {term}, we use the steady state density matrix to express $\delta \hat{\mathcal{F}}_\alpha$ as 
\begin{eqnarray}
\delta \hat{\mathcal{F}}_\alpha = - \bar \beta (\partial_\alpha \hat H - \partial_\alpha \hat Y - Tr(\hat \rho_{ss}  (\partial_\alpha \hat H -\partial_\alpha \hat Y) ) ).
\end{eqnarray}
At equilibrium, $\partial_\alpha \hat Y$ vanishes, such that $\delta \hat{\mathcal{F}}_\alpha$ reduces to the random force operator, $\delta \hat{\mathcal{F}}_\alpha = - \bar \beta (\partial_\alpha \hat H- Tr(\hat \rho_{ss}  \partial_\alpha \hat H))$. Hence, we recover our previous results: the entropy production rate is related to the friction tensor, $\bar T \Delta \dot S_{NA} = \sum_{\alpha \nu} \dot R_{\alpha} \dot R_{\nu} \gamma_{\alpha \nu}$. Out of equilibrium, $\partial_\alpha \hat Y$ does not vanish, such relationship does not hold, and the friction tensor is not positive definite. The entropy production rate, however, remains positive under nonequilibrium conditions.

\section{Application to model systems} \label{sec:model}
In this section, we illustrate the theory discussed above and analyze  thermodynamic quantities for representative model systems. To avoid ambiguities related to a  partitioning between the subsystem and baths, we focus on local thermodynamic quantities, e.g. local population, work, and current. That being said, the calculation of heat and energy do require a partitioning of couplings between the subsystem and baths. For a proper treatment of such cases, see discussions in Ref.\ \onlinecite{PhysRevB.98.134306}.  The numerical simulations are carried out using the HQME method. \cite{ThossHEOM,erpenbeck2019hierarchical,PhysRevB.101.075422} More details of this method can be found in the appendix.

\subsection{The resonant level model} 
The quantum thermodynamics of the resonant level model has been studied in literature. \cite{mishaPRBfriction,PhysRevLett.114.080602,PhysRevB.93.115318} However, most studies are restricted to the equilibrium case, i.e. without any electron current. Here we  study an out-of-equilibrium resonant level model, where a single Fermionic level $d$ (representing, e.g., a level of molecule or a quantum dot) strongly couples to  two macroscopic Fermionic baths:
\begin{eqnarray}
\label{eq:RLMH}
\hat H = E_d (t) \hat d^\dagger \hat d + \sum_{k, \zeta} \epsilon_{k\zeta} \hat c^\dagger_{k\zeta} \hat c_{k\zeta} + \sum_{k,\zeta} V_{k\zeta} ( \hat c^\dagger_{k\zeta} \hat d + \hat d^\dagger \hat c_{k\zeta} ).
\end{eqnarray}
Here $\zeta \in (L, R)$ indicate the left and right leads,  which are described by a continuum of noninteracting Fermionic levels with energies $\epsilon_{k\zeta}$ each. We assume the leads to have the same temperature $k_BT$ but different chemical potentials $\mu_L$ and $\mu_R$ respectively.
We can define the hybridization function $\Gamma_\zeta$ to describe the strength of coupling between $d$ level and the $\zeta$ lead, 
\begin{eqnarray} \label{eq:Gamma_zeta}
\Gamma_{\zeta} (\epsilon) = 2\pi \sum_{k} |V_{k\zeta}|^2 \delta (\epsilon - \epsilon_{k\zeta}) .
\end{eqnarray}
We will apply the wide-band approximation, such that $\Gamma_{\zeta}$ does not depend on $\epsilon$ (and the real part of the self-energy vanishes). The total coupling $\Gamma = \Gamma_L + \Gamma_R$ quantifies the timescale of the overall dynamics. Further, due to external driving, the energy of the $d$ level $E_d (t)$ is time-dependent, described by the following form, 
\begin{eqnarray} \label{eq:E_d}
E_d (t) = \dot E_d t + E_0 ,
\end{eqnarray}
where $E_0$ is the energy level at the starting point (before turning on driving). $\dot{E_d}$ defines the driving speed. To be more explicit, the ratio $\frac{\hbar \dot E_d}{\Gamma^2} $  quantifies slow or fast driving. When $\frac{\hbar \dot E_d}{\Gamma^2} \ll 1 $, we reach the adiabatic limit.

For such a model, the steady state density matrix $\hat \rho_{ss}$ can be obtained analytically. The $\hat Y$ operator in the steady state $\hat \rho_{ss}$ (Eq. (\ref{eq:rho_ss})) equals \cite{wang2013multilayer,PhysRevLett.70.2134,PhysRevB.73.245326,PhysRevLett.99.236808,PhysRevB.75.035302}, 
\begin{eqnarray}
\hat Y =  \sum_{k,\zeta} \mu_\zeta  \hat \psi^\dagger_{k \zeta} \hat \psi_{k\zeta}.
\end{eqnarray}
Here $\hat \psi_{k,\zeta}^\dagger$ is a linear combinations of operators $\hat c_{k\zeta}^\dagger$ and  $\hat d^\dagger$, 
\begin{eqnarray} \label{eq:psi}
\hat \psi_{k\zeta}^\dagger =  \hat c_{k\zeta}^\dagger + V_{k'\zeta'} G(\epsilon_{k\zeta})\left( \hat d^\dagger + \sum_{k',\zeta'}\frac{V_{k'\zeta'}} {\epsilon_{k\zeta} + i\eta - \epsilon_{k'\zeta'}} \hat c_{k'\zeta'}^\dagger \right), 
\end{eqnarray}
and we have defined the retarded Green's function of the dot, 
\begin{eqnarray}
G(\epsilon_{k\zeta})= \frac{1}{\epsilon_{k\zeta} - E_d + i\Gamma/2}.
\end{eqnarray}
The total Hamiltonian in Eq. (\ref{eq:RLMH}) can also be diagonalized by $\hat \psi_{k,\zeta}^\dagger$: 
\begin{eqnarray}
\hat H = \sum_{k,\zeta} \epsilon_{k\zeta} \hat \psi^\dagger_{k \zeta} \hat \psi_{k\zeta}. 
\end{eqnarray}

Using the steady state solution, we can then calculate the population of the dot analytically in the zeroth order of driving, 
\begin{eqnarray} \label{eq:N0}
N^{(0)} = Tr(\hat \rho_{ss} \hat d^\dagger \hat d) = \int \frac{d\epsilon}{2\pi} A(\epsilon) \bar f(\epsilon).
\end{eqnarray}
Here, $A(\epsilon)$ and $\bar{f}$ are spectral function and averaged Fermi distribution, respectively, 
\begin{eqnarray}
A(\epsilon) = \frac{\Gamma}{(\epsilon-E_d)^2 + (\Gamma/2)^2} ,\\
\bar{f}(\epsilon) = \frac{\Gamma_L f^L(\epsilon) + \Gamma_R f^R (\epsilon) }{\Gamma},
\end{eqnarray}
and $f^\zeta (\epsilon) = 1/(1+ \exp(\beta (\epsilon -\mu_\zeta)) )$ is the Fermi function. Further,  the first order nonadiabatic correction to the population is obtained as
\begin{eqnarray} \label{eq:N1}
N^{(1)} = Tr(\hat \rho^{(1)} \hat d^\dagger \hat d) = - \hbar \dot E_d \int \frac{d\epsilon}{4\pi} A^2(\epsilon)  \partial_\epsilon \bar f(\epsilon).
\end{eqnarray}
Here, $\hat \rho^{(1)}$ is given in Eq. (\ref{eq:rho_1}). See the Appendix in Ref.\ \onlinecite{PhysRevB.98.134306} (or Ref.\ \onlinecite{semenov2019transport}) for a derivation of the above result. 

\begin{figure}[htbp] 
   \centering
   \includegraphics[width=4in]{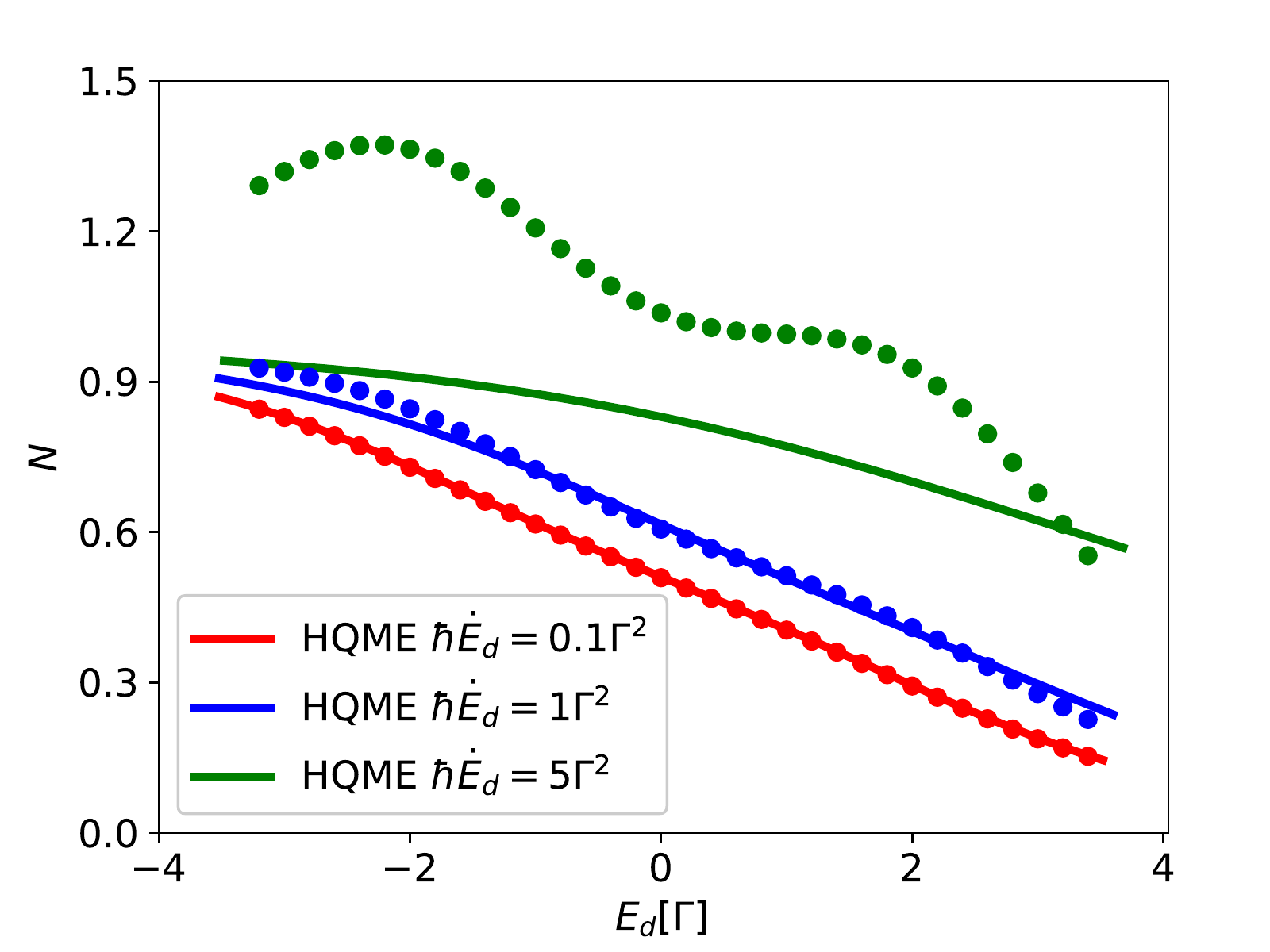} 
   \caption{Population $N$ as a function of $E_d$ from numerical exact results (HQME, lines) and analytical results (dots). Note that $E_d$ is time dependent: $E_d (t) = \dot E_d t + E_0$. We have set the starting point $E_0$ at low enough energy,  such that the HQME results are independent of $E_0$. The analytical results are evaluated up to first order correction in driving rate, i.e. a combination of Eq. (\ref{eq:N0}) and Eq. (\ref{eq:N1})). Note that in the slow driving case $\hbar \dot E_d = 0.1 \Gamma^2$, the first order correction agrees with the numerically exact results. As we increase the driving rates, the first order correction to the population start to deteriorates. In the strongly nonadiabatic regime, our analytical results break down completely and predict unphysical values for the population (greater than 1). The parameters are set to be $kT = \Gamma$, $\mu_L = -\mu_R = 2\Gamma$, $\Gamma_L = \Gamma_R = \frac12 \Gamma$. }
   \label{fig:pop}
\end{figure}

Employing the HQME method, numerically exact results for the 
population as a function of $E_d(t)$ can be obtained,  
 \begin{eqnarray} \label{eq:Nfull}
N = Tr(\hat \rho \hat d^\dagger \hat d) .
\end{eqnarray}
In Fig. \ref{fig:pop}, we plot the populations as a function of $E_d$ obtained from the HQME method and our first order correction (combination of Eq. (\ref{eq:N0}) and Eq. (\ref{eq:N1})) for different driving rates $\dot E_d$. Here $E_d$ is time dependent, $E_d (t) = \dot E_d t + E_0$. We have set the starting point $E_0$ at low enough energy,  such that the HQME results are independent of $E_0$. For a slow driving rate as compared to system dynamics, $\hbar \dot E_d = 0.1 \Gamma^2$, the analytical results agree with the numerically exact result very well. As we increase the driving rate, e.g. $\hbar \dot E_d = 1 \Gamma^2$, the analytical results start to deviate from the HQME results, as the first order correction to the population deteriorates. In the strongly nonadiabatic regime,  $\hbar \dot E_d = 5 \Gamma^2$, the analytical results break down completely and predict unphysical values for the population (greater than 1). Obviously, in the strongly nonadiabatic regime, our approach based on a perturbative treatment of the driving speed is not valid.

To second order in the driving speed, the rate of change in the work is related to the population as follows 
\begin{eqnarray}
\dot W^{(2)} = \dot E_d Tr (\hat \rho^{(1)} \frac{\partial \hat H}{\partial E_d}) = \dot E_d  N^{(1)} = \gamma \dot E_d^2.
\end{eqnarray}
Here, $\gamma$ is the friction coefficient 
\begin{eqnarray} \label{eq:friction}
\gamma = - \hbar \int \frac{d\epsilon}{4\pi} A^2(\epsilon)  \partial_\epsilon \bar f(\epsilon).
\end{eqnarray}
From the numerical simulations, we can quantify friction by the correction to the steady state population divided by the driving speed. 
\begin{eqnarray} \label{eq:friction2}
\gamma =  \frac{N - N^{(0)}}{\dot E_d} .
\end{eqnarray}
For small driving speed, the above equation recovers our definition of friction in (\ref{eq:N1}).

\begin{figure}[htbp] 
   \centering
   \includegraphics[width=4in]{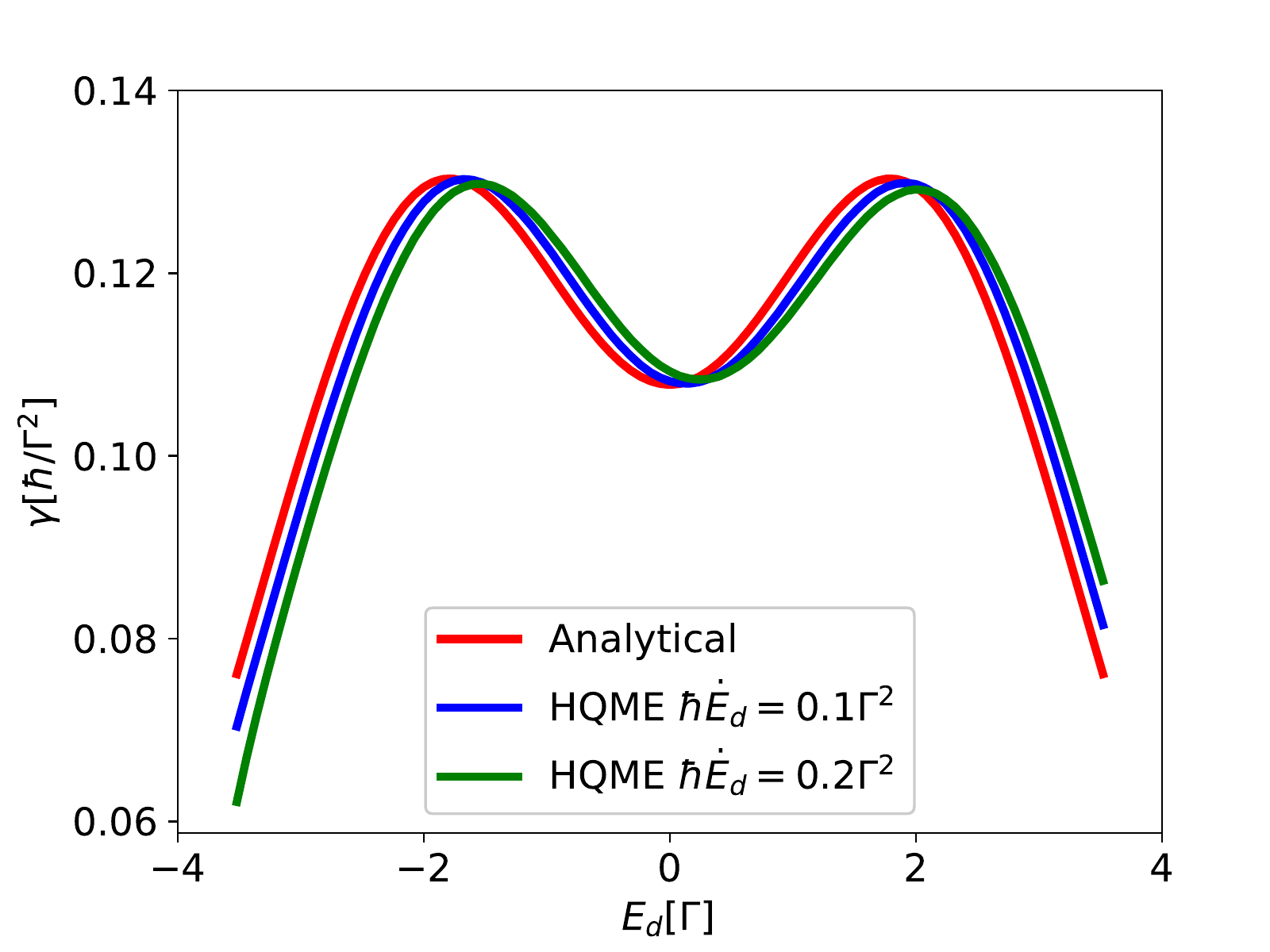} 
   \caption{Friction coefficient $\gamma$ as a function of $E_d$. Analytical results are obtained from Eq. (\ref{eq:friction}) and  HQME results are obtained by Eq. (\ref{eq:friction2}). We note that the friction coefficient exhibits two peaks. The presence of the peaks is due to Fermi resonance ($E_d = \mu_L$ or $E_d = \mu_R$): when the energy of dot level gets close to the chemical potential of the left or right lead, there is a dramatic change in population or work, such that the friction exhibit peaks near the chemical potentials. Again, HQME results agree with analytical analysis in general, with small shift as we increase the driving speed. These shifts are higher orders in driving speed. Note that the analytical result for the friction coefficient (Eq. (\ref{eq:friction})) is independent of $\dot E_d$.  The parameters are set to be $kT = \Gamma$, $\mu_L = -\mu_R = 2\Gamma$, $\Gamma_L = \Gamma_R = \frac12 \Gamma$.}
   \label{fig:friction}
\end{figure}

In Fig. \ref{fig:friction}, we the plot the friction coefficient calculated from  Eq. (\ref{eq:friction}) and the HQME result from Eq. (\ref{eq:friction2}). We note that the friction coefficient exhibits two peaks. The presence of the peaks is due to Fermi resonance: when the energy of the dot level is close to the chemical potential of the left or right lead, there is a dramatic change in population or work, such that the friction coefficient exhibits peaks near the chemical potentials. Again, the HQME results agree with analytical analysis in general, with small shift as we increase the driving speed. These shifts are higher order corrections in the driving speed.

\subsection{Inclusion of electron-electron interactions: the Anderson impurity model}
In Ref.\ \onlinecite{PhysRevLett.119.046001,PhysRevB.98.134306}, we have studied the quantum thermodynamics of the Anderson impurity model at equilibrium, where the el-el interactions give rise to Kondo resonance in thermodynamic quantities. We now analyze such a model out of equilibrium with different chemical potentials from left and right leads:  
\begin{eqnarray} \label{eq:SAIM}
\hat H = E_d (t) \sum_\sigma \hat d^\dagger_\sigma \hat d^\dagger_\sigma + U  \hat d^\dagger_\uparrow \hat d^\dagger_\uparrow \hat d^\dagger_\downarrow \hat d^\dagger_\downarrow + \sum_{k, \zeta, \sigma} \epsilon_{k\zeta} \hat c^\dagger_{k\zeta\sigma} \hat c_{k\zeta\sigma} + \sum_{k,\zeta,\sigma} V_{k\zeta} ( \hat c^\dagger_{k\zeta\sigma} \hat d_\sigma + \hat d^\dagger_\sigma \hat c_{k\zeta\sigma} ).
\end{eqnarray}
Here $U$ is the local Coulomb repulsion energy, and $\sigma=\uparrow,\downarrow$ indicate spin direction. For this interacting model, analytical results are not available. Our analysis below is based on the HQME method. Just as for the resonant level model, we can define hybridization functions as in Eq. (\ref{eq:Gamma_zeta}) and use the wide-band approximation. We will also assume a linear dependence of $E_d$ on time (Eq. (\ref{eq:E_d})).

\begin{figure}[htbp] 
   \centering
   \includegraphics[width=4in]{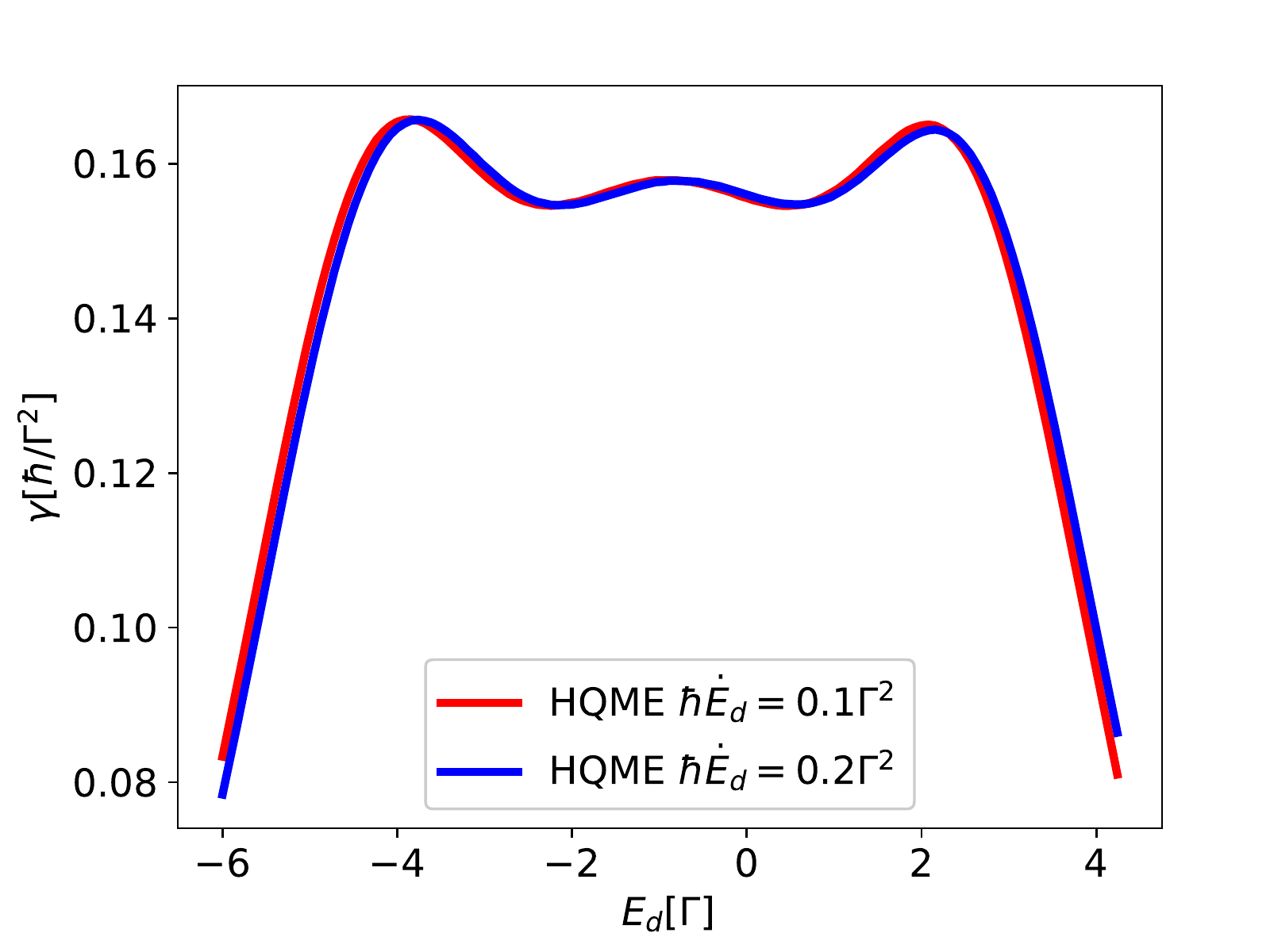} 
   \caption{Friction coefficient $\gamma$ as a function of $E_d$ obtained from the HQME method using Eq. (\ref{eq:friction2}). We note that $\gamma$ exhibits three peaks when the Coulomb repulsion $U$ is non zero. Note that, for the Anderson impurity model, there are effectively two energy levels, $E_d$ and $E_d+U$. The three peaks correspond to resonances where there is a significant change in population: 1) when the first of the two levels starts to approach the lower chemical potential ($E_d + U = \mu_L$), 2) when the last of the two levels start to leave the upper chemical potential ($E_d = \mu_R$), and 3) when the two levels are located exactly between the two chemical potentials ($2E_d + U = \mu_L + \mu_R$). 
The parameters are set to be $kT = \Gamma$, $\mu_L = -\mu_R = 2\Gamma$, $\Gamma_L = \Gamma_R = \frac12 \Gamma$, $U=2\Gamma$.}
   \label{fig:friction_AIM}
\end{figure}

We first analyze the friction coefficient $\gamma$ for the Anderson impurity model based on  the HQME method using Eq. (\ref{eq:friction2}). As shown in Fig. \ref{fig:friction_AIM}, $\gamma$ exhibits three peaks as compared to two peaks in the case of the resonant level model in Fig. \ref{fig:friction}. Note that, for the Anderson impurity model, there are effectively two energy levels, $E_d$ and $E_d+U$. The three peaks correspond to resonances where there is a significant change in population: 1) when the first of the two levels starts to approach the lower chemical potential ($E_d + U = \mu_L$), 2) when the last of the two levels start to leave the upper chemical potential ($E_d = \mu_R$), and 3) when the two levels are located exactly between the two chemical potentials ($2E_d + U = \mu_L + \mu_R$). 

Next, we analyze the nonadiabatic correction to the transport properties of the Anderson impurity model. To this end, we consider the electronic current, which is given by
\begin{eqnarray}
I = -\frac{i}{2\hbar} Tr([ \hat H, \hat N_L - \hat N_R] \hat \rho).
\end{eqnarray}
Here $\hat N_\zeta = \sum_k  \hat c^\dagger_{k\zeta} \hat c_{k\zeta} $ is the number operator for the $\zeta = (L, R)$ lead. 
The steady-state current is obtained using the steady state density matrix, 
\begin{eqnarray}
I^{(0)} = -\frac{i}{2\hbar} Tr([ \hat H, \hat N_L - \hat N_R] \hat \rho_{ss}).
\end{eqnarray}
Just as the definition of friction for the population (or energy) in the above, we can quantify the nonadiabatic correction to the current in the slow driving case by the difference of $I$ and $I^{(0)}$
\begin{eqnarray} \label{eq:gammaI}
\delta I =  \frac{I - I^{(0)}}{\dot E_d} .
\end{eqnarray}

\begin{figure}[htbp] 
   \centering
   \includegraphics[width=4in]{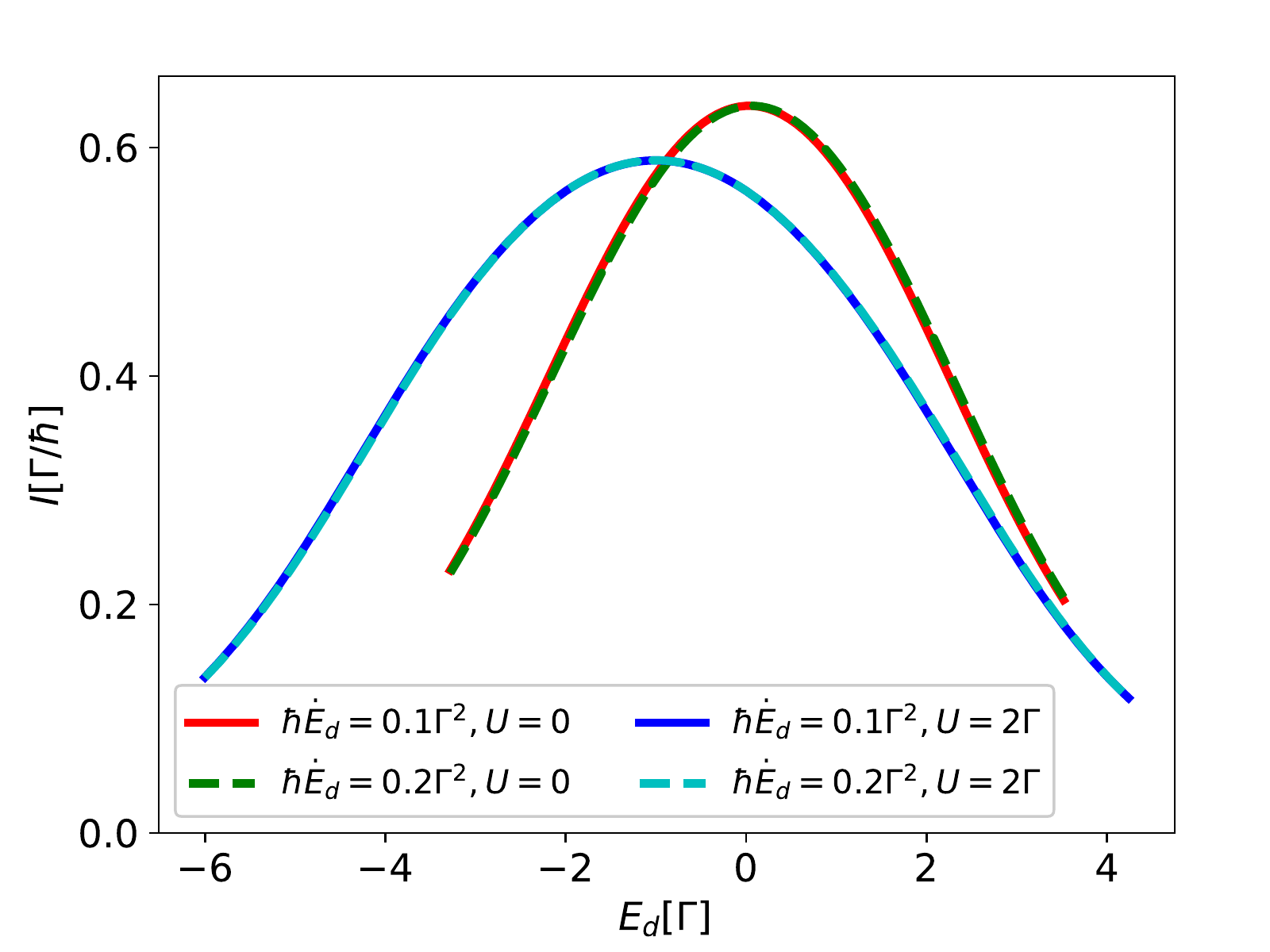} 
   \caption{Electron current $I$ as a function of $E_d$ for different driving speed $\dot E_d$ obtained from the HQME method. We note that the current shows a peak when the two effective dot levels ($E_d$ and $E_d + U$) are located exactly between the two chemical potentials ($2E_d + U = \mu_L + \mu_R$). The peak of the current shifts with the Coulomb repulsion $U$. $kT = \Gamma$, $\mu_L = -\mu_R = 2\Gamma$, $\Gamma_L = \Gamma_R = \frac12 \Gamma$.}
   \label{fig:currentI}
\end{figure}

Before analyzing the nonadiabatic correction to the current, we first consider the electron current itself. Fig. \ref{fig:currentI} shows the current as a function of $E_d$ for different driving speed $\dot E_d$ obtained from the HQME method. We note that the current shows a peak when the two effective dot levels ($E_d$ and $E_d + U$) are located exactly between the two chemical potentials ($2E_d + U = \mu_L + \mu_R$). The peak of the current shifts with the Coulomb repulsion $U$. While the current does not show notable difference for slow driving speeds, the nonadiabatic contribution to the current can reveal more interesting structures (see below).

\begin{figure}[htbp] 
   \centering
   \includegraphics[width=4in]{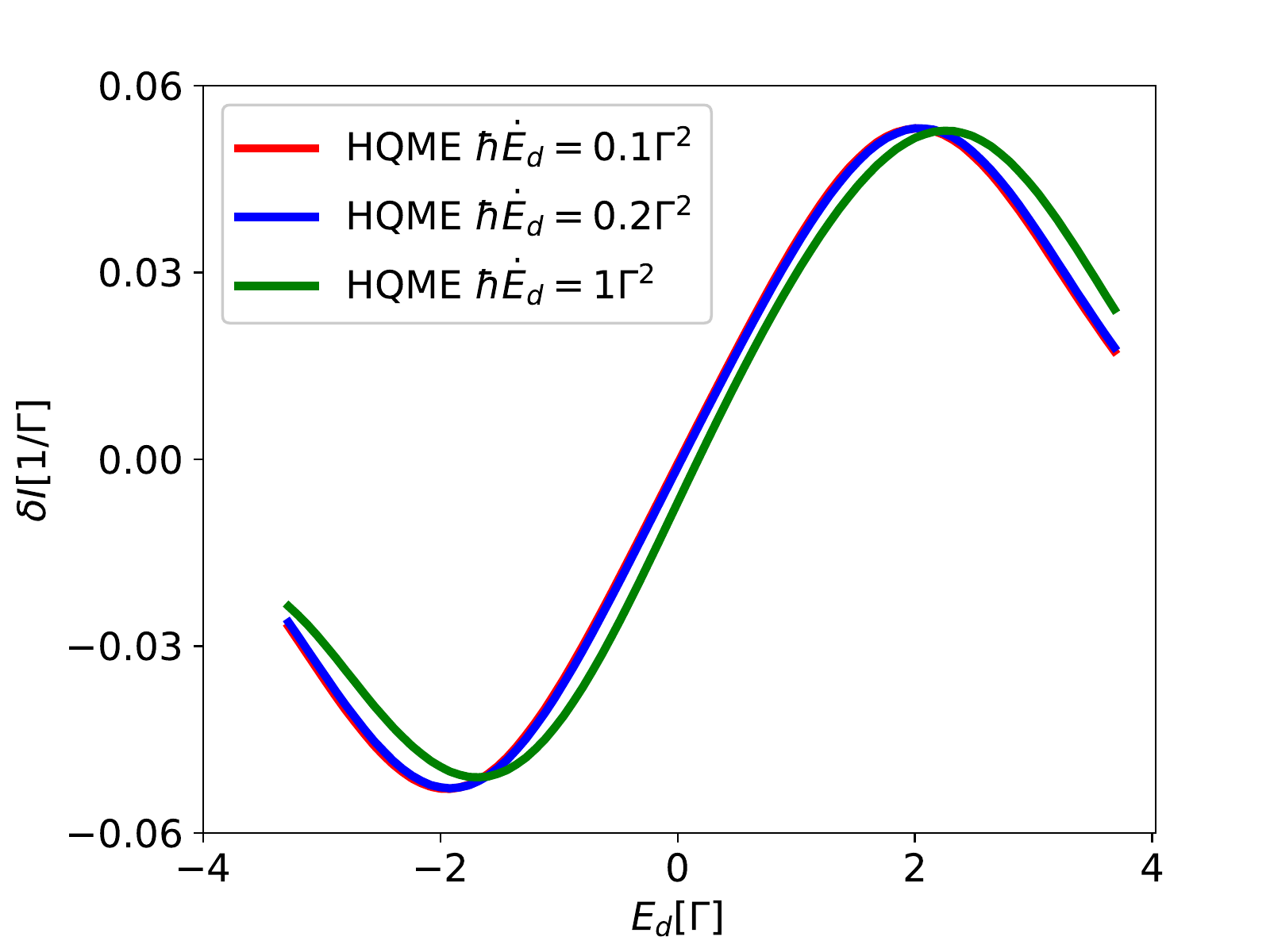} 
   \caption{Nonadiabatic contribution to the current $\delta I$ (Eq. (\ref{eq:gammaI})) as a function of $E_d$ for the Anderson impurity model when $U= 0$. Note that $\delta I$ exhibits a dip near one chemical potential and a peak near the other. This again is due to Fermi resonance. The sign of the current indicates the direction of electron flow. $kT = \Gamma$, $\mu_L = -\mu_R = 2\Gamma$, $\Gamma_L = \Gamma_R = \frac12 \Gamma$, $U=0$.  }
   \label{fig:current}
\end{figure}

We next analyze the nonadiabatic contribution to the current. 
Fig. \ref{fig:current} depicts $\delta I$ (Eq. (\ref{eq:gammaI})) as a function of $E_d$ for the case $U=0$ and different driving speeds. Again, near the chemical potentials, due to Fermi resonance, $\delta I$ exhibits peaks (or dips). The sign of the current indicates the direction of the electron flow. The nonadiabatic contribution to the current $\delta I$ exhibits opposite signs at different chemical potentials, hence a peak near one chemical potential and a dip near the other.

\begin{figure}[htbp] 
   \centering
   \includegraphics[width=4in]{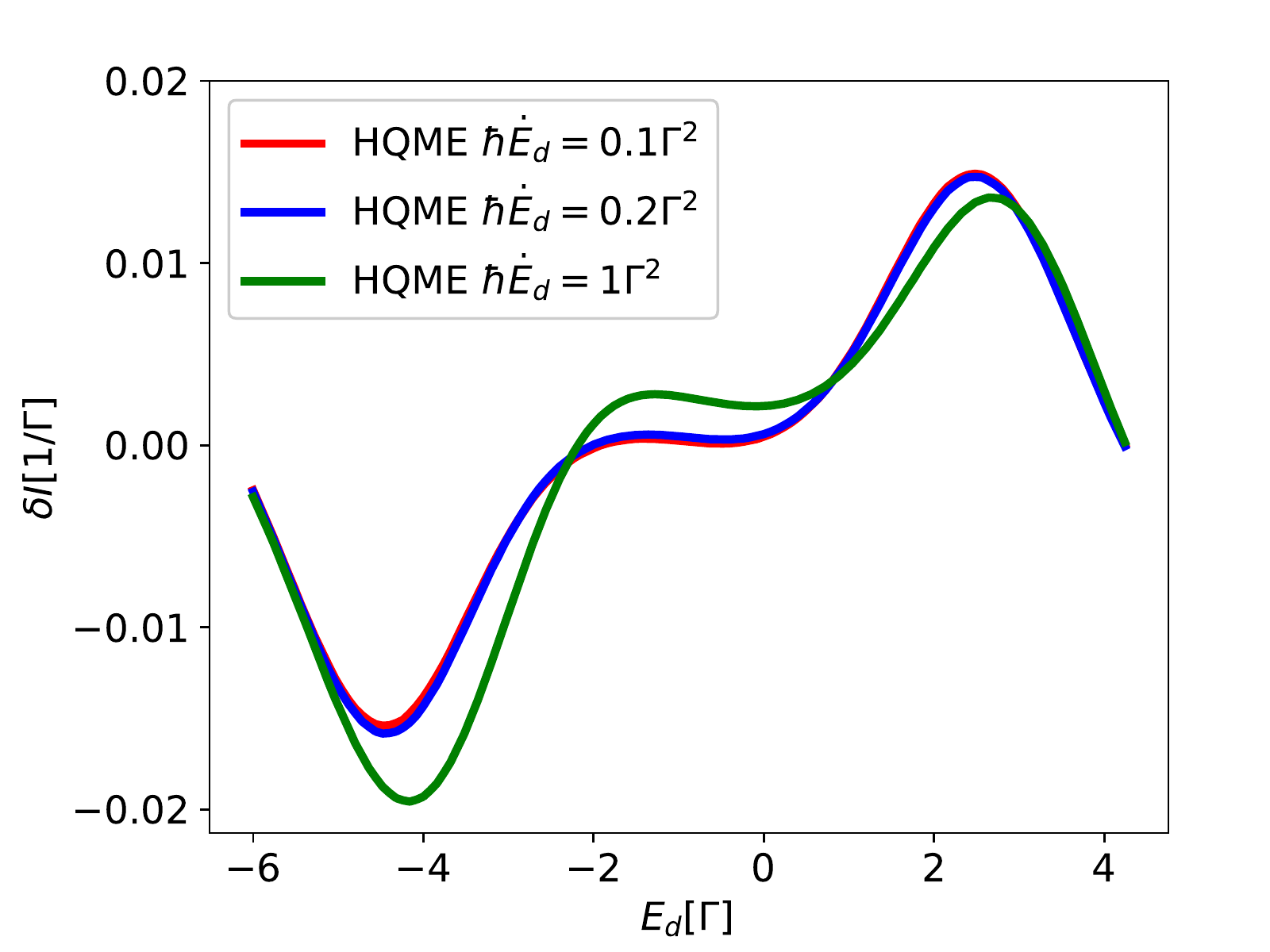} 
   \caption{Nonadiabatic contribution to the current $\delta I$ as a function of $E_d$ for the Anderson impurity model when $U\neq 0$. Note that $\delta I$ exhibits more peaks at energies where the two levels are in resonance with chemical potentials in the leads. Particularly, we see the peaks or dips at $E_d = \mu_L$ ($E_d=-2\Gamma$), $E_d = \mu_R$ ($E_d=2\Gamma$), $E_d + U = \mu_R$ ($E_d=0$), as well as $E_d + U = \mu_L$ ($E_d=-4\Gamma$). $kT = \Gamma$, $\mu_L = -\mu_R = 2\Gamma$, $\Gamma_L = \Gamma_R = \frac12 \Gamma$, $U=2\Gamma$. }
   \label{fig:currentU}
\end{figure}

For the case of $U\neq 0$, when the dot level can be doubly occupied, there is a local Coulomb repulsion between the two electrons with different spins, such that  we have effectively two levels $E_d$ and $E_d + U$ for the dot. In Fig. \ref{fig:currentU}, the nonadiabatic contribution to the current $\delta I$ exhibits more peaks/dips as these two levels are in resonance with chemical potentials in the leads. Specifically, we see peaks or dips at $E_d = \mu_L$ ($E_d=-2\Gamma$), $E_d = \mu_R$ ($E_d=2\Gamma$), $E_d + U = \mu_R$ ($E_d=0$), as well as $E_d + U = \mu_L$ ($E_d=-4\Gamma$). Again, for larger driving speed, we see a slight shift in the position of the peaks/dips. Note that we are not in the Kondo regime. Previously, we have shown that in the limit of strong el-el interactions and low temperature, thermodynamic quantities exhibit Kondo resonance in addition to Fermi resonance at equilibrium.\cite{PhysRevLett.119.046001,PhysRevB.98.134306} Further work addressing the effect of Kondo resonance in thermodynamic quantities under nonequilibrium condition is appropriate. 

\section{conclusions} \label{sec:conclusion}

Within a general framework based on a full density matrix expansion, we 
have formulated the first and second law of thermodynamics for a quantum 
system strongly coupled to two or more baths under nonequilibrium 
conditions and additional external driving. We have quantified the rate 
of entropy production using a Kubo transformed correlation function and 
shown that it remains positive. At equilibrium, our results recover 
previous studies \cite{PhysRevB.98.134306} and the entropy production 
rate can be related to dissipative (frictional) work. The nonequilibrium 
formulation is quite general and can be applied both for Bosonic and 
Fermionic systems. In the present work, we have applied the formalism to 
analyze the resonant level model as well as the Anderson impurity model. 
The nonequilibrium quantum dynamics was simulated using the HQME method, 
which allows a numerically exact solution. The results obtained for the 
Anderson impurity model show that el-el interaction manifests itself as 
Coulomb-blockade signatures in the thermodynamic quantities.

Upon writing this article, we became aware of recent 
work\cite{semenov2019transport}, addressing similar problems using a 
scattering states approach. The work presented here extends this 
important contribution in several ways. We prove that the nonadiabatic 
entropy production is positive in systems carrying a nonequilibrium 
particle and/or heat current. Furthermore, the combination with the HQME 
approach allows the study of interacting problems where the scattering 
states are not analytically available.

\begin{acknowledgements}
This work was supported by the Department of Energy, Photonics at Thermodynamic Limits Energy Frontier Research Center, under Grant No. DE-SC0019140 and a research grant of the German Research Foundation (DFG). 
Furthermore, support by the state of Baden-W\"urttemberg through bwHPC 
and the DFG through Grant No. INST 40/467-1 FUGG is gratefully acknowledged. M.T. thanks Eran Rabani for many insightful discussion on nonequilibrium quantum transport and for kindly hosting his sabbatical stay at the Chemistry Department of the University of California at Berkeley. 
\end{acknowledgements}

\appendix*
\section{Hierarchical Quantum Master Equation (HQME)} \label{sec:hqme}
In the following, we provide some details regarding the numerically exact HQME approach which was used to test our newly developed expansion. The HQME method (also known as hierarchical equation of motion (HEOM) approach) was originally developed in the context of relaxation dynamics\cite{tanimura1989time,tanimura2006stochastic} and later on applied to charge transport\cite{jin2008exact,PhysRevB.88.235426,ThossHEOM}. Here, we closely follow Ref.\ \onlinecite{ThossHEOM}. In contrast to Ref.\ \onlinecite{ThossHEOM}, the HQME approach is described for a time-dependent Anderson impurity model system without vibrational degrees of freedom. 

The derivation of the HQME is based on the system-bath partitioning 
\begin{align}
 \hat{H} = \hat{H}_\tS(t) + \hat{H}_\tSB + \hat{H}_\tB,
\end{align}
where the individual parts are defined according to Eq. \ref{eq:SAIM},
\begin{subequations}
\begin{align}
\hat H_\tS(t) =& E_d (t) \sum_\sigma \hat d^\dagger_\sigma \hat d^\dagger_\sigma + U  \hat d^\dagger_\uparrow \hat d^\dagger_\uparrow \hat d^\dagger_\downarrow \hat d^\dagger_\downarrow ,\\
\hat H_\tSB =& \sum_{k,\zeta,\sigma} V_{k\zeta} ( \hat c^\dagger_{k\zeta\sigma} \hat d_\sigma + \hat d^\dagger_\sigma \hat c_{k\zeta\sigma} ) ,\\
\hat H_\tB =&\sum_{k, \zeta, \sigma} \epsilon_{k\zeta} \hat c^\dagger_{k\zeta\sigma} \hat c_{k\zeta\sigma}.
\end{align}
\end{subequations}

\noindent
Employing a bath interaction picture, the bath coupling operators are defined by
\begin{align}
 \hat b^s_{\zeta \sigma} (t) =& \expo{\ii \hat{H}_\tB t/\hbar} \left( \sum_{k} V_{k\zeta} \hat{c}^s_{k\zeta\sigma} \right) \expo{-\ii \hat{H}_\tB t/ \hbar},
\end{align}
with $s=\pm$, $\hat{c}_{k\zeta\sigma}^{-}\equiv \hat{c}_{k\zeta\sigma}$ and $\hat{c}_{k\zeta\sigma}^{+}\equiv \hat{c}_{k\zeta\sigma}^{\dagger}$. As these operators obey Gaussian statistics, all information about system-bath coupling is encoded in the two-time correlation function of the free bath $C^s_{\zeta,\sigma}(t-\tau)=\avg{ \hat{b}_{\zeta \sigma}^s(t) \hat{b}_{\zeta \sigma}^{\bar s} (\tau) }{\tB}$ where $\bar s \equiv -s$.
Via Fourier transformation
\begin{align}
  C^s_{\zeta,\sigma} (t)=\frac{1}{2 \pi} \int_{-\infty}^\infty \dd \epsilon\, \e^{s \ii \epsilon t/\hbar} \Gamma_{\zeta,\sigma} (\epsilon) f [s (\epsilon-\mu_\zeta)],
\label{eq:C_FT}
\end{align}
$C^s_{\zeta,\sigma}(t)$ is related to the spectral density in the leads $\Gamma_{\zeta,\sigma} (\epsilon)$ and the Fermi-Dirac distribution $f(\epsilon)=\left( \expo{\epsilon/ k_\tB T} +1 \right)^{-1}$.
To derive a closed set of equations of motion within the HQME method, $C^s_{\zeta,\sigma}(t)$ is expressed by a sum over exponentials.\cite{jin2008exact}
To this end, the Fermi distribution is represented by a sum-over-poles scheme employing a Pad\'e decomposition\cite{PhysRevB.75.035123,hu2010communication,hu2011pade} and the spectral density of the leads is assumed to be a single, spin-independent Lorentzian $\Gamma_{\zeta,\sigma} (\epsilon)=\frac12 \frac{\Gamma W^2}{(\epsilon-\mu_\zeta)^2 +W^2}$. The band width $W$ is set to be $10^3$ times larger than $\Gamma$ to effectively describe the leads in the wide-band limit, which implies that the overall molecule-lead coupling strength is independent of energy and symmetric, $\Gamma_\tL = \Gamma_\tR=\frac12 \Gamma$. Thus, the correlation function of the free bath is given by $C^s_{\zeta,\sigma}(t)=\sum_{l=0}^{l_\text{max}} \eta_{\zeta,\sigma,l} \Gamma_\zeta  \e^{-\gamma_{\zeta,\sigma,s,l} t}$. 

The HQMEs are given by
%
\begin{align}
\frac{\partial}{\partial t} \hat{\rho}^{(n)}_{j_n \cdots j_1} =&- \left( \frac{\ii}{\hbar} \hat{ \LL}_\tS(t) + \sum_{m=1}^n \gamma_{j_m} \right) \hat \rho^{(n)}_{j_n \cdots j_1} - \frac{\ii \Gamma}{\hbar} \sum_j \hat \AA^{\bar s}_\sigma \hat \rho^{(n+1)}_{jj_n \cdots j_1} \nonumber\\
&- \ii \sum_{m=1}^n (-)^{n-m} \hat \CC_{j_m} \hat \rho^{(n-1)}_{j_n \cdots j_{m+1}j_{m-1} \cdots j_1},
\label{eq:HQME}
\end{align}
%
%
with the multi-index $j=(\zeta,\sigma,s,l)$ and $\hat{\tilde \LL}_\tS(t) \hat O = [ \hat{\tilde H}_\tS(t) , \hat O ]$.
Here, $\hat \rho^{(0)} \equiv \hat \rho$ represents the reduced density matrix and $\hat \rho^{(n)}_{j_n \cdots j_1}$ $(n>0)$ denote auxiliary density matrices, which describe bath-related observables such as, e.g., the current
\begin{align}
\av{\hat{I}_\zeta (t)}= \ii \frac{e \Gamma}{2\hbar} \sum_{\sigma,l} \text{Tr}_\tS \left\{ \hat d_\sigma  \hat{\rho}_{\zeta,\sigma,+,l}^{(1)}(t) - \text{h.c.} \right\}.
\end{align}
The superoperators $\hat \AA$ and $\hat \CC$ read
\begin{subequations}
\begin{align}
 \hat \AA^{\bar s}_\sigma \hat \rho^{(n)} =& \hat{d}^{\bar s}_\sigma \hat \rho^{(n)} + (-)^n \hat \rho^{(n)} \hat{d}^{\bar s}_\sigma , \\
\hat \CC_{\zeta,\sigma, s,l} \hat \rho^{(n)}  =& \eta_{\zeta,\sigma,l} \hat{d}^s_\sigma   \hat \rho^{(n)} - (-)^n \eta^{*} _{\zeta,\sigma,l} \hat \rho^{(n)} \hat{d}^s_\sigma .
\end{align}
\end{subequations}
According to system-bath interaction, the superoperator $\hat \AA$ ($\hat \CC$) couples the $n$th-level of the hierachy to the $(n+1)$th ($(n-1)$th) level. The importance of the auxiliary density operators is estimated by assigning them the following importance values,\cite{PhysRevB.88.235426}
\begin{align}
	 \mathcal{I} \left(\hat \rho^{(n)}_{j_n \cdots j_1}\right) =  \left|\left(\prod\limits_{m=1}^{n-1 }\frac{\Gamma/(2\hbar)}{\sum\limits_{a\in\{1..{m}\}}\hspace{-.2cm} \text{Re}\left[\omega_{j_{a}}\right]}\right) \left( \prod\limits_{m=1}^{n} \frac{\eta_{j_{m}}}{\text{Re}\left[\omega_{j_{m}}\right]} \right)\right|
\end{align}
In the calculations presented in this paper, the results are quantitatively converged for truncations of the hierarchy at level $n=4$, neglecting auxiliary density operators having an importance value smaller $10^{-9}$.

\end{document}